\documentclass[letter]{ieice}

\usepackage[dvips]{graphicx}
\newenvironment{numberedlist}
{\begin{list}{\makebox[20pt]{\hss(\arabic{itemno})\enspace}}
             {\usecounter{itemno}\labelwidth 20pt}}{\end{list}}

\newcounter{itemno}

\newcounter{itemno1}

\newcounter{itemno2}

\newcounter{exno}

\newcounter{defno}

\newenvironment{defn}{\refstepcounter{defno}\medskip \noindent {\bf
Definition \thedefno.\ }}{\medskip}

\newcommand{\sep}{\;\vert\;}

\newcommand{\oprove}{\vdash\kern-.6em\lower.7ex\hbox{$\scriptstyle O$}\,}

\newcommand{\pderivation}{{\cal P}\kern -.1em\hbox{\rm -derivation}}
\newcommand{\pderivationl}{{\cal P}\kern -.1em\hbox{\em -derivation}}
\newcommand{\pderivable}{{\cal P}\kern -.1em\hbox{\rm -derivable}}
\newcommand{\pderivablel}{{\cal P}\kern -.1em\hbox{\em -derivable}}
\newcommand{\pderivations}{{\cal P}\kern -.1em\hbox{\rm -derivations}}
\newcommand{\pderivability}{{\cal P}\kern -.1em\hbox{\rm -derivability}}

\newcommand{\all}{\forall}
\newcommand{\some}{\exists}

\newcommand{\ie}{{\em i.e.}}

\newsavebox{\lpartfig}
\newsavebox{\rpartfig}

\newenvironment{exmple}{
 \begingroup \begin{tabbing} \hspace{2em}\= \hspace{3em}\= \hspace{3em}\=
\hspace{3em}\= \hspace{3em}\= \hspace{3em}\= \kill}{
 \end{tabbing}\endgroup}

\newcommand{\lb}{\langle}
\newcommand{\rb}{\rangle}

\newcommand{\intp}{intp_o}
  
\newcommand{\add}{\oplus} 

     {\\* \hspace*{\fill} \end{trivlist}}

\field{}

\title{Priority, Cut, If-Then-Else and Exception Handling in Logic Programming}
\authorlist{
\authorentry{Keehang KWON}{m}{labelA}
}
\affiliate[labelA]{The author is a professor of Computer Eng., DongA University. email:khkwon@dau.ac.kr}
\received{2003}{1}{1}
\revised{2003}{1}{1}
\finalreceived{2003}{1}{1}
\renewcommand{\add}{\bigtriangledown} 
\newcommand{\addp}{\bigtriangledown^*} 
\newcommand{\adcp}{\bigtriangleup^*} 
\begin{document}
\maketitle
\begin{summary}
  One of the long-standing problems on logic programming is to express {\it priority}-related
  operations -- default reasoning, if-then-else, cut,
exception handling, etc --  in a  high-level way.
 We argue that this problem can be  solved by adopting computability logic and 
   $prioritized$ sequential-disjunctive goal formulas of the form 
 $G_0 \addp G_1$  where  $G_0, G_1$ are
goals.   These goals have the following intended semantics:
 sequentially $choose$ the first successful goal $G_i$, while discarding the rest,
  and execute $G_i$ where $i (= 0\ {\rm or}\ 1)$.  
 These goals thus allow us to specify a task $G_0$ with the failure-handling (exception handling) routine
 $G_1$. The operator $\addp$ can also be seen  as a logic-equivalent of the $if$-$then$-$else$ statement in imperative
 language.
 We also discuss prioritized sequential-conjunction clauses which are {\it dual} of sequential-disjunctive goals.
\end{summary}

\newcommand{\muprolog}{{Prolog$^{\addp,\adcp}$}}

\renewcommand{\intp}{ex} 


\section{Introduction}\label{sec:intro}

One of the major  problems on logic programming is to treat  priority-related 
primitives in a  {\it logical} way.
The progress  of logic programming  has enriched the theory of Horn clauses with
higher-order programming, mutual exclusion,  etc.
 Nevertheless some related issues -- priority, default reasoning, cut, exception handling -- 
 could not be dealt with elegantly.

    In this paper, we propose a unified solution to these problems. It involves the direct employment of computability logic (CoL) \cite{Jap03}
 to allow for   goals with exception handling capability. CoL presented a class of
  sequential disjunctive  goals $G_0 \add  G_1$,
 where $G_0, G_1$ are goals.  It is capable of  $updating$  the knowledge/query from $G_0$ to $G_1$.
 It naturally supports exception handling in the sense that if $G_0$ fails, $G_1$ will be attempted.  On the other hand, it 
 does not support priority between $G_0$ and $G_1$: there is no preference between $G_0$ and $G_1$ if both
 goals are solvable.
 
 To overcome this problem, we present  prioritized sequential disjunctive (PSD) goals $G_0\addp G_1$. 
Executing this goal with respect to a program $D$ -- $\intp(D,G_0 \addp  G_1)$ --
 has the following intended semantics: 
 sequentially choose, while discarding the rest,  the first successful  one  between
\[ \intp(D, G_0), \intp(D, G_1). \]

    An
illustration of this  aspect is provided by the following definition of the
relation $sort(X,Y)$ which holds if $Y$ is a sorted list  of $X$:

\begin{exmple}
$      sort(X,Y)$ ${\rm :-}$ \> \hspace{5em}      $heapsort(X,Y)\ \addp\ quicksort(X,Y)$ 
\end{exmple}
\noindent
The body of the definition above contains a PSD goal, denoted by $\addp$.
 As a particular example, solving the query $sort([3,100,40,2],Y)$ would result in selecting and 
executing the first goal $heapsort([3,100,40,2],Y)$.  If the heapsort module is available in the program, then the given goal 
will
succeed, producing solutions for $Y$. If the execution fails for some reason(heapsort not defined,
stack overflow, etc), the machine tries the plan B, \ie, the quicksort module.

The operator $\addp$ is, in fact, {\it indispensable} to logic programming, as it is a
logic-programming equivalent of the $if$-$then$-$else$ in
imperative languages. To see this, consider the following example:
\begin{exmple}
$      max(X,Y,Z)$ ${\rm :-}$ \> \hspace{6em}      $(X > Y \land Z = X)\ \addp\ (Z = Y)$.
\end{exmple}
Of course, we can specify PSD goals using cut
in Prolog, but it is well-known that cuts
complicates the declarative meaning of the program \cite{Bratko}.
Our language makes it possible to formulate 
{\it mutually exclusive} goals in a high-level way.
The class of sequential-disjunctive goals is a high-level abstraction for the cut
predicate.

We'd like to point out that the class of PSD-goals has a dual notion. That is, prioritized sequential-choice
is possible at the clause level via prioritized sequential-conjunctive clauses of the form $D_0\adcp D_1$. For example, $max$
can be rewritten as follows:
\begin{exmple}
  $      (max(X,Y,X)$ ${\rm :-}$ \> \hspace{6em}   $X > Y)\ \adcp$ \\
  $      (max(X,Y,Y))$. \>
\end{exmple}
\noindent In this setting, the machine tries the first clause. If it fails, then it tries the next
and so on.
This aspect -- which we call {\it negative} exception handling (or negative if-then-else) --
was discussed  in the context of functional languages\cite{Kfunc}.

Our $\adcp$-clauses can be used to encode {\it default} rules. For example, unless we have a proof
that a bird has a broken wing, we can infer that it can fly. 

\begin{exmple}
$      (nofly(X)$ ${\rm :-}$ \> \hspace{4em}      $brokenwing(X) \land   bird(X))$ $\adcp$ \\
$      (fly(X)$ ${\rm :-}$ \> \hspace{4em}      $ bird(X)$). \\
\end{exmple}
\noindent This is related to priority logic programming and defeasible logic programming.

\section{The Language}\label{sec:logic}

The language, called \muprolog,  is a version of Horn clauses
 with PSD goals and PSC clauses. 
It is described
by $G$- and $D$-formulas given by the syntax rules below:
\begin{exmple}
\>$G ::=$ \>   $A \sep  G \land  G \sep    \some x\ G \sep  G \addp G $ \\   \\
\>$D ::=$ \>  $A  \sep G \supset A\ \sep \all x\ D \sep D \land D \sep D\adcp D$\\
\end{exmple}
\noindent
\newcommand{\sync}{up}
\newcommand{\async}{down}

In the rules above,  
$A$  represents an atomic formula.
A $D$-formula  is called a  Horn
 clause with sequential choice.

\newcommand{\bc}{bc}

We will  present a machine's strategy, which we call {\it prioritized}-proof.
The idea is that we associate a pmove with $D_0\adcp D_1$ where a {\it pmove} is a bit $0$ or $1$.
 During execution, the  pmove  0(1) is recorded, $if D_0 (D_1)$ is used. A {\it prun} is a finite sequence of pmoves.
We propose a priority-based derivation of the form $(\cal D, \rho)$ which associates a prun $\rho$ with the
derivation $\cal D$.  A prun places a priority on
the derivation $\cal D$.  Our scheme is based on the popular lexicographical ordering .
 That is, a prun  000$\ldots$ means the highest and 111$\ldots$  the lowest priority.
For example,  suppose there are two derivations $(\mathcal{D}_1,01)$ and $(\mathcal{D}_2,11)$ for 
a goal.  Then it expresses that  the former has higher-priority than the latter.

 Resolving conflicts based on lexicographical ordering on pruns might look
complicated but is in fact quite simple and leads to the above semantics.  

To be specific, the way the machine  operates in fact depends on the top-level 
constructor in the expression,  a property known as
uniform provability\cite{Mil89jlp,MNPS91,KK07}. 
 Note that execution  alternates between 
two phases: the goal-reduction phase 
and the backchaining phase. 
In  the goal-reduction phase (denoted by $\intp(D,G)$), the machine tries to solve a goal $G$ from
a clause $D$  by simplifying $G$. The rule (9) -- (12) 
are related to  this phase.
If $G$ becomes an atom, the machine switches to the backchaining mode. 
This is encoded in the rule (8). 
In the backchaining mode (denoted by $bc(D_1,D,A)$), the machine tries 
to solve an atomic goal $A$ 
by first reducing a Horn clause $D_1$ to simpler forms and then 
backchaining on the resulting 
 clause. 

\begin{defn}\label{def:semantics}
Let $G$ be a goal and let $D$ be a program.
Then the notion of   executing $\lb D,G\rb$ -- $\intp(D,G)$ -- 
 is defined as follows:

\begin{numberedlist}

\item  $\bc(A,D,A)$. \%  SUCC rule.

\item    $\bc((G_0\supset A),D,A)$ if 
  $\intp(D, G_0)$. \% APPLY rule
  
\item    $\bc(D_0\land D_1,D,A)$ if   $\bc(D_0,D,A)$. \% AND-L-1 rule 

\item    $\bc(D_0\land D_1,D,A)$ if    $\bc(D_1,D,A)$.  \% AND-L-2 rule 

\item    $\bc(D_0\adcp D_1,D,A)$ if   $\bc(D_0,D,A)$. \% CAPITAL-L rule

\item    $\bc(D_0\adcp D_1,D,A)$ if    $\bc(D_1,D,A)$, provided $\bc(D_0,D,A)$ is a
  failure.  \% SWITCH-L rule

\item    $\bc(\all x D_1,D,A)$ if   $\bc([t/x]D_1,D, A)$. \% ALL-L rule

\item    $\intp(D,A)$ if   $\bc(D,D, A)$. \% BACKCHAIN rule


\item  $\intp(D,G_0 \land G_1)$  if $\intp(D,G_0)$  and
  $\intp(D,G_1)$. \% AND-R rule

\item $\intp(D,\exists x G_0)$  if $\intp(D,[t/x]G_0)$. \% SOME-R rule

\item $\intp(D,G_0\addp G_1)$ if
$\intp(D, G_0)$. \% CAPITAL-R rule
   
\item $\intp(D,G_0\addp G_1)$ if 
$\intp(D, G_1)$, provided $\intp(D, G_0)$ is a
  failure.  \% SWITCH-R rule. This goal behaves as a goal with exception handling.
\end{numberedlist}
\end{defn}

\noindent  
In the above rules,  the rule  (11)-(12) are novel features.
  To be specific, this goal first attempts to execute $G_0$.
 If it succeeds, then do nothing (and do not leave any choice point for $G_1$
). If it fails, then $G_1$ is attempted. The rule (5)-(6) can be treated similarly.

\section{Examples }\label{sec:modules}

As an  example, let us consider the following database which contains the today's flight
information for major airlines such as Panam and Delta airlines.

\begin{exmple}
\% panam(source, destination, dp\_time, ar\_time) \\
\% delta(source, destination, dp\_time, ar\_time) \\
$panam(paris, nice, 9:40, 10:50)$\\
$\vdots$\\
$panam(nice, london, 9:45, 10:10)$\\
$delta(paris, nice, 8:40, 09:35)$\\
$\vdots$\\
$delta(paris, london, 9:24, 09:50)$\\
\end{exmple}
\noindent Consider a goal $\some dt \some at\ panam(paris,london,dt,at) \addp \some dt \some at\ delta(paris,london,dt,at)$. This goal expresses the task of finding whether the user has a 
flight in Panam to fly from paris to london today. Since there is no Panam flight, 
  the system now tries Delta.  Since Delta has a flight, 
the system produces the departure and arrival
time of the flight of the Delta airline.

\section{Conclusion}\label{sec:conc}

In this paper, we have considered an extension to Prolog with  
PSD  goals and PSC clauses. These goals and clauses are 
particularly useful for specifying priority, cut, if-then-else and
exception handling in Prolog, making Prolog
more versatile.

Our semantics of $\addp$ in Section 2 
is based on prioritized-derivations that agree with the specified priority.
A bottom-up prioritized-derivation admits the SWITCH-L rule to $D_i$ if there are no other derivations via $D_0,\ldots,D_{i-1}$. Similarly, the SWITCH-R rule for $G_i$.
Our semantics of $\addp$  can be seen as a conservative extension of $\add$ of CoL \cite{Jap03,Jap08}, as
all the derivations of CoL without using $\addp$ remain to have the equal priority.
For example,  solving a query $\some x (p(x)\add q(x))$ from $p(a) \land q(b)$ would produce
two legal answers, ie, $x = a$ or $x = b$  in our semantics.
If we switch from $\add$ to $\addp$, the machine produces only one legal answer $x = a$.


\bibliographystyle{ieicetr}

\begin{thebibliography}{1}




\bibitem{Bratko}
I.~Bratko,   ``Prolog:programming for AI '',
 addpison Wesley, 2001 (3rd edition). 


\bibitem{HM94}
J.~Hodas and D.~Miller, ``Logic Programming in a Fragment of Intuitionistic Linear Logic'', Information and Computation, vol.110, pp.327--365, 1994.


\bibitem{Jap03}
G.~Japaridze, ``Introduction to computability logic'', Annals  of Pure and
 Applied  Logic, vol.123, pp.1--99, 2003.

\bibitem{Jap08}
G.~Japaridze,   ``Sequential operators in computability logic'',
 Information and Computation, vol.206, No.12, pp.1443-1475, 2008.  


\bibitem{Kfunc}
  K.Kwon and D.Kang, ``Extending functional languages with high-level exception handling'',
  http://arXiv.1709.04619. 

\bibitem{KK07}
E.~Komendantskaya and V.~Komendantsky, ``On uniform proof-theoretical operational semantics for logic programming'',  In J.-Y. Beziau and A.Costa-Leite, editors, Perspectives on Universal Logic, pages 379--394. Polimetrica Publisher, 2007.

\bibitem{Mil89jlp}
D.~Miller, ``A logical analysis of modules in logic programming'', Journal of
  Logic Programming, vol.6, pp.79--108, 1989.

\bibitem{MNPS91}
D.~Miller, G.~Nadathur, F.~Pfenning, and A.~Scedrov, ``Uniform proofs as a
  foundation for logic programming'', Annals of Pure and Applied Logic, vol.51,
  pp.125--157, 1991.

 


\end{thebibliography}


\end{document}